\documentclass[10pt,a4paper]{article}
\usepackage{amssymb}
\usepackage{amsmath}

\begin{document}

\author{Ph. H\"{a}gler$^1$, A. Mukherjee$^2$, A. Sch\"{a}fer$^3$}

\title{\begin{flushleft}
       \mbox{\normalsize DO-TH 03/15}
       \end{flushleft}
       \vskip 20pt
       Quark Orbital Angular Momentum in the Wandzura--Wilczek Approximation}


\maketitle

\begin{center}
$^1$Department of Physics and Astronomy\\
Vrije Universiteit,
De Boelelaan 1081,
1081 HV Amsterdam, The Netherlands\\
phone: +31 20 4447906
fax: +31 20 4447999
email: haegler@p2h.de\\
\vspace{1mm}
$^2$Institut f\"{u}r Physik\\
Universit\"{a}t Dortmund, D-44221 Dortmund, Germany\\
\vspace{1mm}
$^3$Institut f\"{u}r Theoretische Physik\\
Universit\"{a}t Regensburg, D-93040 Regensburg, Germany
\end{center}


\begin{abstract}
We show that quark orbital angular momentum is directly related to
off-forward correlation functions which include intrinsic transverse
momentum corresponding to a derivative with respect to the transverse
coordinates. Its possible contribution to scattering processes is therefore
of higher twist and vanishes in the forward limit. The relation of OAM to
other twist 2 and 3 distributions known in the literature is derived and
formalized by an unintegrated sum rule.
\end{abstract}

\section{Introduction}

It is well known that orbital angular momentum of quarks and gluons can give
an important contribution to the total spin of the proton, according to the
sumrule 
\begin{equation}
\frac{1}{2}=\frac{1}{2}\Delta q+L_{q}+\Delta g+L_{g},  \label{sumrule1}
\end{equation}
where $\Delta q,L_{q},\Delta g$ and $L_{g}$ are the first moments of the
corresponding quark spin $\Delta q(x)$, quark OAM $L_{q}(x)$, gluon spin $%
\Delta g(x)$ and gluon OAM distribution $L_{g}(x)$. Further on one knows
since long how to measure polarized distribution functions and much work was
invested into reliable estimates of their leading moments. Still, despite
enormous theoretical and experimental efforts only $\Delta q$ is 
known to be rather small, \cite{Abe98,Ackerstaff:1999ey,Goto2000}, 
and the largest value given in the literature is of the order of 
$\Delta q(Q^{2}\simeq 1$GeV$^{2})\simeq 0.30$ in the $\overline{\mbox{MS}}$ scheme, 
while our knowledge of
$\Delta g$ is extremely limited \cite{Bar02,Airapetian:1999ib}. Finally, for a long time it was
not known at all how to access the quark and gluon OAM pieces
experimentally. With the advent of studies of off-forward scattering
processes like DVCS, $L_{q}$ and $L_{g}$ came for the first time within
reach, although not directly, but via the sumrule \cite{Ji97} 
\begin{equation}
\frac{1}{2}\int dxx\left( H_{q,g}(x,\xi )+E_{q,g}(x,\xi )\right) =J_{q,g},
\label{sumrule2}
\end{equation}
where $H(x,\xi )$ and $E(x,\xi )$ are off-forward distributions in the limit
of vanishing momentum transfer squared $\Delta ^{2}\rightarrow 0$. This
sumrule allows to extract e.g. the quark OAM contribution, assuming
knowledge of the quark spin $\Delta q,$ via 
\begin{equation}
L_{q}=J_{q}-\frac{1}{2}\Delta q.  \label{sumruleInt}
\end{equation}
While an experimental determination of $H_{q,g}(x,\xi ,\Delta ^{2})$ and $%
E_{q,g}(x,\xi ,\Delta ^{2})$ with an accuracy which allows to estimate $J_{q}
$ and $J_{g}$ with interesting precision is still a long way to go, eq. (%
\ref{sumrule2}) was used recently to calculate $J_{q}$ on the lattice \cite
{Goeckeler2003},\cite{Haegler2003}.

In ref.\cite{Pen00} a sum rule has been presented relating quark OAM to the
second moment of the twist 3 off-forward distribution $G_{3}(x)$ (now $%
G_{2}(x)$). Our notation is that GPDs depending only on $x$ denote the
GPDs taken in the forward limit, e.g. $G_i(x)=G_i(x,\xi=0 ,\Delta ^{2}=0)$.
Including the new $G_{4}(x)$ distribution \cite{Pol02}, this
sumrule reads 
\begin{equation}
\int dxx\left[ G_{2}(x)-2G_{4}(x)\right] =-L_{q}.  \label{sumrule3}
\end{equation}
It is useful to notice that the identification eq.(\ref{sumrule3}) has been
made using the sumrule eq.(\ref{sumrule2}). On the level of \emph{%
distributions}, Hoodbhoy et al. \cite{Ji99} showed that the sumrule eq.(\ref
{sumrule2}) is valid for higher moments as well and therefore in the forward
limit 
\begin{equation}
\frac{1}{2}x\left( q(x)+E(x)\right) =J_{q}(x),  \label{sumrule2a}
\end{equation}
where it has been used that $H(x)$ is equal to the unpolarized quark
distribution $q(x)$. Furthermore Hoodbhoy et al. defined quark spin and OAM
distributions (in terms of inverse Melin-transformed higher moments) which
are evidently interrelated by 
\begin{equation}
L_{q}(x)=J_{q}(x)-\frac{1}{2}\Delta q(x).  \label{sumrule6}
\end{equation}
Combining eq.(\ref{sumrule2a}) and eq.(\ref{sumrule6}), it is possible to
determine the quark OAM distribution directly from measurable quantities
using the sumrule \cite{Ji99} 
\begin{equation}
L_{q}(x)=\frac{1}{2}x\left( q(x)+E(x)\right) -\frac{1}{2}\Delta q(x).
\label{sumrule7}
\end{equation}
In this study we show that the parton model definition of the quark OAM
distribution can be related to a certain off-forward matrix element as soon
as intrinsic transverse momenta are taken into account. Using the equation
of motion, we derive an unintegrated sum rule in the Wandzura-Wilczek
approximation which relates $L_{q}(x)$ to some of the known twist 2 and 3
off-forward distributions.

\section{Quark OAM, intrinsic transverse momentum and off-forward correlators}

\subsection{Rewriting Quark OAM}

We begin by recalling the definition of the quark OAM distribution in the
light cone gauge according to \cite{Jaf98} 
\begin{equation}
f_{L_{q}}(x)=\int dx^{-}e^{ix\frac{P^{+}x^{-}}{2}}\frac{\left\langle
P\right| \int d^{2}x_{\perp }\psi _{+}^{\dagger }(x_{\perp })i\left(
x^{1}\partial _{2}-x^{2}\partial _{1}\right) \psi _{+}(x_{\perp
}+x^{-})\left| P\right\rangle }{4\pi \left( \int d^{2}x_{\perp }\right) },
\label{OAM01}
\end{equation}
with $\psi _{+}=1/2 \gamma^- \gamma^+ \psi$ and where we have replaced 
the residual gauge covariant derivative by the
partial derivative, $\mathcal{D}=\partial-ig\mathcal{A}\rightarrow \partial$. 
If the boundary conditions can be fixed so that the $A_\perp$-fields vanish
at infinity, then the residual gauge field $\mathcal{A}$ would be exactly equal to zero.
However in the light cone gauge, non-vanishing gluon fields at the boundary 
can give rise to topological effects, as is discussed in \cite{Harindranath1993}. 
How this could effect our calculation lies beyond the scope of this presentation.
First we observe that one easily runs into problems
due to the explicit factors and the integration of $x_{\perp }$ in case that
one takes the definition eq.(\ref{OAM01}) literally, see below. Our plan is
now to reformulate eq.(\ref{OAM01}) in order to circumvent this from the
beginning by incorporating an additional transverse vector which gives us
some handle on the transverse direction. For this purpose we introduce the
function $f_{L_{q}}(x,\Delta _{\bot })$ defined as 
\begin{equation}
f_{L_{q}}(x,\Delta _{\bot })=\int \frac{dx^{-}d^{2}x_{\perp }}{4\pi }e^{ix%
\frac{P^{+}x^{-}}{2}}\left\langle P^{\prime }\right| \psi _{+}^{\dagger
}(x_{\perp })i\left( x^{1}\partial _{2}-x^{2}\partial _{1}\right) \psi
_{+}(x_{\perp }+x^{-})\left| P\right\rangle ,  \label{OAM02}
\end{equation}
which differs from eq.(\ref{OAM01}) mainly in that the matrix element is
slightly off-forward in the transverse direction, i.e. $P^{\prime }=\bar{P}%
+\Delta _{\bot }/2,P=\bar{P}-\Delta _{\bot }/2$. The function $%
f_{L_{q}}(x,\Delta _{\bot })$ is well defined when acting on a test function 
$T(\Delta _{\bot })$, and we therefore define the quark OAM as 
\begin{equation}
f_{L_{q}}(x)\equiv \frac{1}{(2\pi )^{2}T(0)}\int d^{2}\Delta _{\bot
}T(\Delta _{\bot })f_{L_{q}}(x,\Delta _{\bot }).  \label{OAMconv}
\end{equation}
We see that the original definition eq.(\ref{OAM01}) can be reproduced by
choosing $T(\Delta _{\bot })=\delta ^{2}(\Delta _{\bot })$. Shifting the
fields and writing the factor $x_{\perp }$ as derivative with respect to $%
\Delta _{\bot }$, we find after partial integration 
\begin{eqnarray}
f_{L_{q}}(x) &=&\frac{1}{(2\pi )^{2}T(0)}\int d^{2}\Delta _{\bot }\frac{%
dx^{-}}{4\pi }e^{ix\frac{P^{+}x^{-}}{2}}(2\pi )^{2}\delta ^{2}(\Delta _{\bot
}) \nonumber\\
&&\epsilon _{jk}\partial _{\Delta_\bot j}\left\{ T(\Delta _{\bot
})\left\langle P^{\prime }\right| \psi _{+}^{\dagger }(0_{\perp })\left.
\partial_{x_\bot k}\psi _{+}(x_{\perp }+x^{-})\right| _{x_\bot =0}\left|
P\right\rangle \right\} \label{OAM03a} .
\end{eqnarray}
where the antisymmetric $\epsilon_{jk}=1$ for $j=1,k=2$ and 0 for $j=k$.
Since we get no contribution in case that the derivative acts on the
testfunction (e.g. if $T(\Delta _{\bot})$ is symmetric in $\Delta _{\bot}$), 
we end up with 
\begin{equation}
f_{L_{q}}(x)=\int \frac{dx^{-}}{4\pi }e^{ix\frac{P^{+}x^{-}}{2}}\epsilon
_{jk}\partial _{\Delta_\bot j}\left\{ \left\langle P^{\prime }\right| \psi
_{+}^{\dagger }(0_{\perp })\left. \partial _{x_\bot k}\psi _{+}(x_{\perp
}+x^{-})\right| _{x_\bot =0}\left| P\right\rangle \right\} _{\Delta_\bot =0}.
\label{OAM03}
\end{equation}
The choice $T(\Delta _{\bot })=\delta ^{2}(\Delta _{\bot })$ would result in
an integral over the square of the delta function, which is not well
defined. This is an indication for the difficulties associated with a naive
use of (\ref{OAM01}).

Another way to circumvent these potential problems is to introduce wave
packets $\left| \phi\right\rangle$, e.g. similar to the
discussion in \cite{Burkardt00}. We then would start with the forward
distribution $f(x)$ and make the integration over the transverse coordinate 
$x_{\bot }$ as in eq.(\ref{OAM01}) more explicit by writing 
\begin{equation}
f(x)=\int d^{2}x_{\bot }g(x,x_{\bot })
\equiv\int d^{2}x_{\bot }
\left\langle P\right|x_\perp\hat{O}(x,x_{\bot })\left| P\right\rangle. \label{unint1}
\end{equation}
The function $g(x,x_{\bot })$ can now be treated in exactly the same way as the
impact parameter dependent distribution in \cite{Burkardt00}, and by replacing 
the momentum eigenstates in eq.(\ref{unint1}) by wave packets, we find
\begin{eqnarray}
f(x) &=&
\frac{1}{\left| N\right| ^{2}}\int d^{2}\Delta _{\bot }\delta ^{2}(\Delta
_{\bot })\int d^{2}\bar{P}_{\bot }\phi ^{*}(\bar{P}_{\bot }+\frac{\Delta
_{\bot }}{2})\phi (\bar{P}_{\bot }-\frac{\Delta _{\bot }}{2})
\nonumber \\
&& i\partial _{\Delta_\bot}
\langle \bar{P}_{\bot }+\frac{\Delta
_{\bot }}{2}|\hat{O}(x,x_{\bot }=0) | \bar{P}_{\bot }-\frac{\Delta
_{\bot }}{2}\rangle
\label{wavep1}
\end{eqnarray}
where $N$ is normalization of the wave packet.
This has to be compared with the corresponding result using the testfunction
(see the structure of eq.(\ref{OAM03a}) ), 
\begin{equation}
f(x)=\frac{1}{T(0)}\int d^{2}\Delta _{\bot }\delta ^{2}(\Delta _{\bot
})T(\Delta _{\bot })
i\partial _{\Delta_\bot}\langle\bar{P}_{\bot }+\frac{\Delta
_{\bot }}{2}|\hat{O}(x,x_{\bot }=0)| \bar{P}_{\bot }-\frac{\Delta
_{\bot }}{2}\rangle
\label{testf1}
\end{equation}
From eqs. (\ref{wavep1}) and (\ref{testf1}) we find that $T(\Delta _{\bot })$
plays a role similar to $\int d^{2}\bar{P}_{\bot }\phi ^{*}(\bar{P}_{\bot }+%
\frac{\Delta _{\bot }}{2})\phi (\bar{P}_{\bot }-\frac{\Delta _{\bot }}{2})$.

Orbital angular momentum distribution functions similar to eq. (\ref{OAM01})
have been used in refs.\cite{Harindranath1998} and \cite{Haegler1998} to
calculate their respective evolution equations. A study of the evolution of
off-forward correlators and the sumrule eq.(\ref{sumrule1}) for dressed
quark states in light-front pQCD can be found in \cite{Mukherjee2002}.

\subsection{Quark OAM in terms of proton wave functions}

We now reexamine the above considerations using proton wave functions.
Starting from the decomposition of the proton state $\left| P\right\rangle $
in terms of proton wave functions \cite{Bro01} (for the overlap
representation of GPDs see also \cite{Diehl2001a})
\begin{eqnarray}
\left| P\right\rangle  &=&\sum\limits_{n}\int \left[ \prod_{i=1}^{n}\frac{%
d^{2}k_{i\perp }dx_{i}}{\sqrt{x_{i}}2(2\pi )^{3}}\right] \delta \left(
1-\sum_{i=1}^{n}x_{i}\right) \delta ^{(2)}\left( \sum_{i=1}^{n}k_{i\perp
}\right)   \nonumber \\
&&\left| \left\{ x_{i}P^{+},x_{i}P^{+}+k_{i\perp },\lambda _{i}\right\}
_{i=1..n}\right\rangle \Psi _{n}(x_{i},k_{i\perp },\lambda _{i},\lambda ),
\label{fspace}
\end{eqnarray}
where $\Psi _{n}(x_{i},k_{i\perp },\lambda _{i},\lambda )$ is the n-particle
fock state wave function of the proton. The normalization is given by 
\[
\left\langle p^{+},p_{\perp },\lambda ^{\prime }\right| \left.
k^{+},k_{\perp },\lambda \right\rangle =2p^{+}(2\pi )^{3}\delta _{\lambda
\lambda ^{\prime }}\delta (p^{+}-k^{+})\delta ^{(2)}(p_{\perp }-k_{\perp }).
\]
We choose 
\begin{eqnarray}
P^{+} &=&P^{0}+P^{3},P^{-}=P^{0}-P^{3},  \nonumber \\
P^{\prime } &=&\bar{P}+\Delta /2,P=\bar{P}-\Delta /2,(1+\xi )P^{\prime
+}=(1-\xi )P^{+},  \label{kin1}
\end{eqnarray}
but work most of the time in the limit $\xi=0$. 

Inserting eq.(\ref{fspace}) in eq.(\ref{OAM02}) we end up with 
\begin{eqnarray}
f_{L_{q}}^{WF}(x,\Delta _{\bot }) &=&\frac{1}{2}\sum\limits_{n}\sum%
\limits_{a=1}^{n}\int \left[ \prod\limits_{i=1}^{n}\frac{dx_{i}d^{2}q_{i\bot
}}{2(2\pi )^{3}}\right] \delta \left( x_{a}-x\right) \delta \left(
1-\sum_{i=1}^{n}x_{i}\right)   \nonumber \\
&&\delta ^{(2)}\left( \sum_{i=1}^{n}q_{i\perp }\right) \int d^{2}x_{\bot
}e^{-i\Delta _{\bot }\cdot x_{\bot }}\left\{ x_{\bot }\times
(q_{a}-x_{a}\Delta _{\bot }/2)\right\} _{3}  \nonumber \\
&&\Psi _{n}^{\dagger }(x_{i},\tilde{q}_{i\perp })\Psi _{n}(x_{i},q_{i\perp
}),  \label{OAMWF}
\end{eqnarray}
where 
\[
\tilde{q}_{a\perp }=q_{a\perp }+(1-x_{a})\Delta _{\bot },\tilde{q}_{i\neq
a\perp }=q_{i\perp }-x_{i}\Delta _{\bot }.
\]
We see immediately that the rhs of eq.(\ref{OAMWF}) vanishes if we take the
naive limit $\Delta _{\bot }\rightarrow 0$, which would reproduce the
definition eq.(\ref{OAM01}). Using instead the definition eq.(\ref{OAMconv}%
), rewriting $x_{\bot }$ as a derivative and performing a partial
integration gives 
\begin{eqnarray}
f_{L_{q}}^{WF}(x) &=&\frac{1}{T(0)}\frac{1}{2}\sum\limits_{n}\sum%
\limits_{a=1}^{n}\int \left[ \prod\limits_{i=1}^{n}\frac{dx_{i}d^{2}q_{i\bot
}}{2(2\pi )^{3}}\right] \delta \left( x_{a}-x\right) \delta \left(
1-\sum_{i=1}^{n}x_{i}\right)   \nonumber \\
&&\delta ^{(2)}\left( \sum_{i=1}^{n}q_{i\perp }\right) i\left\{ T(0)\left(
\left\{ q_{a}\times \partial _{\Delta }\right\} _{3}\Psi _{n}^{\dagger }(%
x_{i},\tilde{q}_{i\perp })\Psi _{n}(x_{i},q_{i\perp })\right) \left|
_{\Delta _{\bot }=0}\right. \right.   \nonumber \\
&&\left. +\left( \Psi _{n}^{\dagger }(x_{i},\tilde{q}_{i\perp })\Psi
_{n}(x_{i},q_{i\perp })\left\{ q_{a}\times \partial _{\Delta }\right\}
_{3}T(\Delta _{\bot })\right) \left| _{\Delta _{\bot }=0}\right. \right\} ,
\label{OAMWF2}
\end{eqnarray}
where we have dropped all terms $\varpropto \Delta _{\bot }$. Since the
second term in (\ref{OAMWF2}) vanishes, we end up with 
\begin{eqnarray}
f_{L_{q}}^{WF}(x) &=&\frac{1}{2}\sum\limits_{n}\sum\limits_{a=1}^{n}\int
\left[ \prod\limits_{i=1}^{n}\frac{dx_{i}d^{2}q_{i\bot }}{2(2\pi )^{3}}%
\right] \delta \left( x_{a}-x\right) \delta \left(
1-\sum_{i=1}^{n}x_{i}\right)   \nonumber \\
&&\delta ^{(2)}\left( \sum_{i=1}^{n}q_{i\perp }\right) i\left( \left\{
q_{a}\times \partial _{\Delta }\right\} _{3}\Psi _{n}^{\dagger }(x_{i},
\tilde{q}_{i\perp })\Psi _{n}(x_{i},q_{i\perp })\right) \left| _{\Delta
_{\bot }=0}\right. ,  \label{OAMWF3}
\end{eqnarray}
which is the quark OAM in terms of proton wave functions. This can be
written as 
\begin{equation}
f_{L_{q}}^{WF}(x)=\int \frac{dx^{-}}{4\pi }e^{ix\frac{P^{+}x^{-}}{2}%
}\epsilon _{\bot jk}\partial _{\Delta_\bot j}\left( \left\langle P^{\prime
}\right| \psi _{+}^{\dagger }(x_{\perp })\partial _{x_\perp k}\psi
_{+}(x_{\perp }+x^{-})\left| P\right\rangle \right) \left| _{\Delta _{\bot
},x_{\bot }=0}\right. ,  \label{OAMWF4}
\end{equation}
which is just equal to eq.(\ref{OAM03}).

\subsection{Off-forward correlators and intrinsic transverse momenta}

First we introduce the slightly more general kinematics 
\begin{eqnarray*}
n^{2} &=&p^{2}=0,n\cdot p=1, \\
P^{^{\prime }} &=&(1-\xi )p+(1+\xi )\frac{\bar{M}^{2}}{2}n+\frac{1}{2}\Delta
_{\bot }, \\
P &=&(1+\xi )p+(1-\xi )\frac{\bar{M}^{2}}{2}n-\frac{1}{2}\Delta _{\bot }, \\
\bar{M}^{2} &=&M^{2}-\frac{1}{4}\Delta ^{2}.
\end{eqnarray*}
(In terms of light cone coordinates one can choose 
\begin{eqnarray*}
p^{+} &=&\bar{P}^{+},n^{-}=\frac{2}{\bar{P}^{+}}=\frac{2}{p^{+}}, \\
p^{-} &=&n^{+}=0.)
\end{eqnarray*}
Now let us consider the generic off-forward correlator 
\begin{equation}
\int \frac{d\lambda d^{2}x_{\bot }}{(2\pi )^{3}}e^{ix\lambda -ik_{\bot
}\cdot x_{\bot }}\left\langle P^{\prime }\right| \bar{\psi}(-\frac{\lambda }{%
2}n,-\frac{x_{\bot }}{2})\gamma ^{\mu } \mathcal{U}\psi (\frac{\lambda }{2}n,\frac{%
x_{\bot }}{2})\left| P\right\rangle .  \label{corr1}
\end{equation}
In order to get a gauge invariant correlator, we included a link operator 
$\mathcal{U}$ which runs in particular along the transverse
direction connecting the
points $(-\lambda/2 \,n,-x_{\bot }/2)$ and 
$(\lambda/2 \,n,x_{\bot }/2)$. 
For recent discussions of these transverse gauge links and their
implications see e.g. \cite{Mulders2003,Ji2003}.
Neglecting the intrinsic transverse momentum $k_{\bot }$ in a given 
hard scattering amplitude involving eq.(\ref{corr1}) allows for a 
direct integration of eq.(\ref{corr1}%
) over $k_{\bot }$, and in this case we end up with an expression which is
local in the transverse direction, $x_{\bot }=0$, 
\begin{equation}
\int \frac{d\lambda }{2\pi }e^{ix\lambda }\left\langle P^{\prime }\right| 
\bar{\psi}(-\frac{\lambda }{2}n)\gamma ^{\mu }\psi (\frac{\lambda }{2}%
n)\left| P\right\rangle .  \label{corr1b}
\end{equation}
Such correlations functions have been parametrized in terms of twist 2 and 3
distributions, see e.g.\cite{Ji97,Pen00}. Taking into account terms linear
in the intrinsic transverse momentum $k^{\nu _{\bot }}$ and performing the
integration leads to a correlator with one derivative \cite{Ani00} 
\begin{equation}
\int \frac{d\lambda }{2\pi }e^{ix\lambda }\left\langle P^{\prime }\right| 
\bar{\psi}(-\frac{\lambda }{2}n,-\frac{x_{\bot }}{2})\gamma ^{\mu }%
\overleftrightarrow{\partial ^{\nu _{\bot }}}\psi (\frac{\lambda }{2}n,\frac{%
x_{\bot }}{2})\left| P\right\rangle _{x_{\bot }=0},  \label{corr3}
\end{equation}
where $\overleftrightarrow{\partial }=1/2(\overrightarrow{\partial }-%
\overleftarrow{\partial })$. 
The partial derivative, when acting on the link operator $\mathcal{U}$ in eq.(\ref{corr1})
leads to terms with explicit transverse gluon
operators $A_\perp$. These additional contributions will be neglected in our approximation.
In any case, there are no transverse links left in eqs.(\ref{corr1b},\ref{corr3}) since the
relevant operators are local in transverse direction.
For the following it is important to observe that the correlator eq.(\ref{corr3})
must be proportional to the components of the only remaining transverse
vector, the momentum transfer $\Delta _{\bot }$. We concentrate now on the
part which is proportional to the combination $\epsilon ^{\nu _{\bot }\sigma
_{\bot }}\Delta _{\sigma _{\bot }}$. Then it is possible to parametrize eq.(%
\ref{corr3}) by 
\begin{eqnarray}
\int \frac{d\lambda }{2\pi }e^{ix\lambda }\left\langle P^{\prime }\right| 
\bar{\psi}(-\frac{\lambda }{2}n,-\frac{x_{\bot }}{2})\not{n}%
\overleftrightarrow{\partial ^{\nu _{\bot }}}\psi (\frac{\lambda }{2}n,\frac{%
x_{\bot }}{2})\left| P\right\rangle _{x_{\bot }=0} 
&=&\frac{\epsilon ^{\nu _{\bot }\Delta _{\bot }np}}{2}\bar{U}(P^{\prime
},S^{\prime })\not{n}\gamma _{5}U(P,S)L_{q}(x,\xi ,\Delta ^{2})
\nonumber \\
&+&(\propto \Delta^{\nu_\perp})+(\propto \xi)
\label{corr4}
\end{eqnarray}
where we have already multiplied with $n_{\mu }$. The last two terms on the rhs of
eq.(\ref{corr4}) indicate the presence of contributions proportional to 
$\Delta^{\nu_\perp}$ and $\xi$ which vanish after taking the derivative $\epsilon _{\nu _{\bot }\sigma _{\bot }}\partial _{\Delta
}^{\sigma _{\bot }}$ (see below) and the forward limit.
Of course one can write down other terms for the parametrization in eq.(\ref{corr4})
which only implicitely give rise to a factor $\epsilon ^{\nu _{\bot }\sigma
_{\bot }}\Delta _{\sigma _{\bot }}$. Using (generalized) Gordon identities (see discussion
below eq.(\ref{PARA1})) these contributions can be, however, 
reduced to the term on the rhs of eq.(\ref{corr4}), 
up to the indicated terms which are irrelevant for our calculation.
We now rewrite the lhs and
obtain 
\begin{equation}
\int \frac{dx^{-}}{2\pi }e^{i\frac{P^{+}x^{-}}{2}(x+\xi )}\left( \frac{i}{2}%
\Delta ^{\nu _{\bot }}+\partial _{x}^{\nu _{\bot }}\right) \left\langle
P^{\prime }\right| \psi _{+}^{\dagger }(0)\psi _{+}(x_{\perp }+x^{-})\left|
P\right\rangle \left| _{x_{\bot }=0}\right.   \label{lhs1}
\end{equation}
The function $L_{q}$ in eq.(\ref{corr4}) can then be extracted by taking the
derivative $\epsilon _{\nu _{\bot }\sigma _{\bot }}\partial _{\Delta
}^{\sigma _{\bot }}$ of both sides. For the rhs of (\ref{corr4}) we get in
the forward limit $2L_{q}(x)=2L_{q}(x,\xi =0,\Delta ^{2}=0)$, and the lhs is
given by 
\begin{equation}
\int \frac{dx^{-}}{2\pi }e^{i\frac{P^{+}x^{-}}{2}(x+\xi )}\epsilon _{\nu
_{\bot }\sigma _{\bot }}\partial _{\Delta }^{\sigma _{\bot }}\left(
\left\langle P^{\prime }\right| \psi _{+}^{\dagger }(0)\partial _{x}^{\nu
_{\bot }}\psi _{+}(x_{\perp }+x^{-})\left| P\right\rangle \right) \left|
_{x_{\bot }=0}\right.   \label{lhs2}
\end{equation}
Comparing eq.(\ref{OAMWF4}) and eq.(\ref{lhs2}), we see that in the forward
limit we have the identification 
\begin{equation}
L_{q}(x)=f_{L_{q}}(x).  \label{ident1}
\end{equation}
Thus the off-forward correlator (\ref{corr3}) is directly related to the
quark OAM distribution of the proton.

\subsection{Relation to other twist 2 and 3 off-forward distributions}

Since the correlator eq.(\ref{corr3}) involves a transverse derivative, it
corresponds to (kinematical) twist 3. Similar to the investigations in \cite
{Ani00,Ani01} we apply now the identity 
\begin{eqnarray}
\bar{\psi}(x_{2})\gamma ^{[\mu }\overleftrightarrow{\partial ^{\nu ]}}\psi
(x_{1}) &=&-i\epsilon ^{\mu \nu \alpha \beta }\bar{\psi}(x_{2})\gamma 
\underleftrightarrow{_{\alpha }\partial _{\beta }}\gamma _{5}\psi (x_{1}),
\label{EOM1} \\
\epsilon _{0123} &=&+1,
\end{eqnarray}
coming from the equations of motion, 
where $\underleftrightarrow{\partial }=1/2(\overrightarrow{\partial }+%
\overleftarrow{\partial })$ and where $[\cdots ]$ stays for the
anti-symmetric combination of the indices, to the correlator eq.(\ref{corr4}%
). We contract again with $n_{\mu }$ and choose $\nu =\nu _{\bot }$. This
leads to the following relation 
\begin{eqnarray}
&&\int \frac{d\lambda }{2\pi }e^{ix\lambda }\left\langle P^{\prime }\right| 
\bar{\psi}(-\frac{\lambda }{2}n,-\frac{x_{\bot }}{2})\not{n}%
\overleftrightarrow{\partial ^{\nu _{\bot }}}\psi (\frac{\lambda }{2}n,\frac{%
x_{\bot }}{2})\left| P\right\rangle _{x_{\bot }=0}  \nonumber \\
&=&\int \frac{d\lambda }{2\pi }e^{ix\lambda }\left\langle P^{\prime }\right| 
\bar{\psi}(-\frac{\lambda }{2}n,-\frac{x_{\bot }}{2})\gamma ^{\nu _{\bot
}}n\cdot \overleftrightarrow{\partial }\psi (\frac{\lambda }{2}n,\frac{%
x_{\bot }}{2})\left| P\right\rangle _{x_{\bot }=0}  \nonumber \\
&&-i\epsilon ^{n\nu _{\bot }\alpha \beta }\int \frac{d\lambda }{2\pi }%
e^{ix\lambda }\left\langle P^{\prime }\right| \bar{\psi}(-\frac{\lambda }{2}%
n,-\frac{x_{\bot }}{2})\gamma _{\alpha }\underleftrightarrow{\partial
_{\beta }}\gamma _{5}\psi (\frac{\lambda }{2}n,\frac{x_{\bot }}{2})\left|
P\right\rangle _{x_{\bot }=0}.  \label{EOM2}
\end{eqnarray}
Using 
\begin{equation}
\epsilon ^{n\nu _{\bot }\alpha \beta }\gamma _{\alpha }\underleftrightarrow{%
\partial _{\beta }}=\epsilon ^{n\nu _{\bot }p\beta _{\bot }}\not{n}%
\underleftrightarrow{\partial _{\beta _{\bot }}}+\epsilon ^{n\nu _{\bot
}\alpha _{\bot }p}\gamma _{\alpha _{\bot }}n\cdot \underleftrightarrow{%
\partial },  \label{exp1}
\end{equation}
the third line in eq. (\ref{EOM2}) can be rewritten 
\begin{eqnarray}
&&i\epsilon ^{n\nu _{\bot }p\beta _{\bot }}\int \frac{d\lambda }{2\pi }%
e^{ix\lambda }\left\langle P^{\prime }\right| \bar{\psi}(-\frac{\lambda }{2}%
n,-\frac{x_{\bot }}{2})\not{n}\underleftrightarrow{\partial _{\beta _{\bot }}%
}\gamma _{5}\psi (\frac{\lambda }{2}n,\frac{x_{\bot }}{2})\left|
P\right\rangle _{x_{\bot }=0}  \nonumber \\
&&+i\epsilon ^{n\nu _{\bot }\alpha _{\bot }p}\int \frac{d\lambda }{2\pi }%
e^{ix\lambda }\left\langle P^{\prime }\right| \bar{\psi}(-\frac{\lambda }{2}%
n,-\frac{x_{\bot }}{2})\gamma _{\alpha _{\bot }}n\cdot \underleftrightarrow{%
\partial }\gamma _{5}\psi (\frac{\lambda }{2}n,\frac{x_{\bot }}{2})\left|
P\right\rangle _{x_{\bot }=0}.  \label{EOM3}
\end{eqnarray}
Using partial integration (when possible) and translations, one can show
that the three different derivatives occurring in eq. (\ref{EOM2}) and eq. (%
\ref{EOM3}) can be substituted by 
\begin{eqnarray}
n\cdot \overleftrightarrow{\partial } &\rightarrow &-ix,  \nonumber \\
\underleftrightarrow{\partial _{\beta _{\bot }}} &\rightarrow &\frac{i}{2}%
\Delta _{\beta _{\bot }},  \nonumber \\
n\cdot \underleftrightarrow{\partial } &\rightarrow &\frac{i}{2}n\cdot
\Delta =-i\xi .  \label{SUB1}
\end{eqnarray}
Following this we get for the rhs of eq. (\ref{EOM2}) 
\begin{eqnarray}
&&-ix\int \frac{d\lambda }{2\pi }e^{ix\lambda }\left\langle P^{\prime
}\right| \bar{\psi}(-\frac{\lambda }{2}n,-\frac{x_{\bot }}{2})\gamma ^{\nu
_{\bot }}\psi (\frac{\lambda }{2}n,\frac{x_{\bot }}{2})\left| P\right\rangle
_{x_{\bot }=0}  \nonumber \\
&&-i\epsilon ^{n\nu _{\bot }p\beta _{\bot }}(\frac{i}{2}\Delta _{\beta
_{\bot }})\int \frac{d\lambda }{2\pi }e^{ix\lambda }\left\langle P^{\prime
}\right| \bar{\psi}(-\frac{\lambda }{2}n,-\frac{x_{\bot }}{2})\not{n}\gamma
_{5}\psi (\frac{\lambda }{2}n,\frac{x_{\bot }}{2})\left| P\right\rangle
_{x_{\bot }=0}  \nonumber \\
&&-i\epsilon ^{n\nu _{\bot }\alpha _{\bot }p}(-i\xi )\int \frac{d\lambda }{%
2\pi }e^{ix\lambda }\left\langle P^{\prime }\right| \bar{\psi}(-\frac{%
\lambda }{2}n,-\frac{x_{\bot }}{2})\gamma _{\alpha _{\bot }}\gamma _{5}\psi (%
\frac{\lambda }{2}n,\frac{x_{\bot }}{2})\left| P\right\rangle _{x_{\bot }=0}.
\label{EOM4}
\end{eqnarray}
The correlators in eq. (\ref{EOM4}) can be completely parametrized in terms
of the known twist 2 and 3 off-forward distributions in the
WW-approximation, see \cite{Pol02}. This gives 
\begin{eqnarray}
&&\bar{U}(P^{\prime },S^{\prime })\left\{ -ix\left[ \left( H+E+G_{2}\right)
\gamma ^{\nu _{\bot }}-G_{4}i\epsilon ^{\nu _{\bot }\alpha _{\bot }np}\Delta
_{\alpha _{\bot }}\not{n}\gamma _{5}\right] \right.  \nonumber \\
&&-i\epsilon ^{n\nu _{\bot }p\beta _{\bot }}(\frac{i}{2}\Delta _{\beta
_{\bot }})\left[ \tilde{H}\not{n}\gamma _{5}+\tilde{E}\frac{n\cdot \Delta }{%
2M}\gamma _{5}\right]  \nonumber \\
&&-i\epsilon ^{n\nu _{\bot }\alpha _{\bot }p}(-i\xi )\left[ (\tilde{H}+%
\tilde{G}_{2})\gamma _{\alpha _{\bot }}\gamma _{5}+(\tilde{E}+\tilde{G}_{1})%
\frac{\Delta _{\alpha _{\bot }}}{2M}\gamma _{5}\right.  \nonumber \\
&&\left. \left. +\tilde{G}_{3}\Delta _{\alpha _{\bot }}\not{n}\gamma
_{5}\right] \right\} U(P,S)+(\propto \Delta ^{\nu _{\bot }}),  \label{PARA1}
\end{eqnarray}
where all GPDs are functions of $x,\xi $ and $\Delta $, and where we do not
show the terms proportional to $\Delta ^{\nu _{\bot }}$,
because they are not directly related to the $L_q$-term in eq.(\ref{corr4}).
Using some Gordon-identities \cite{Diehl2001} we have with our kinematics
and in the limit $\xi \rightarrow 0$ 
\[
\epsilon ^{\nu _{\bot }\beta _{\bot }np}\Delta _{\beta _{\bot }}\bar{U}%
(P^{\prime },S^{\prime })\not{n}\gamma _{5}U(P,S)=-2i\bar{U}(P^{\prime
},S^{\prime })\gamma ^{\nu _{\bot }}U(P,S). 
\]
Taking this together with eq.(\ref{corr4}) we end up with the following
sumrule between the distribution $L_{q}(x)$ from and the distribution
functions from eq. (\ref{PARA1}) in the forward limit, 
\begin{equation}
L_{q}(x)=x\left( q(x)+E(x)+G_{2}(x)-2G_{4}(x)\right) -\Delta q(x).
\label{sumrule5}
\end{equation}
where $L_{q}(x)$ is, according to the upper analysis, identified as
(forward) quark OAM in the parton model.

\subsection{Comparing with the integrated G$_{2}$-sumrule}

For reasons of comparison let us now recalculate the integrated sumrule.
Following ref. \cite{Pol01} (eq.50), we have due to equations of motions 
\begin{eqnarray}
&&\int dxx\int \frac{d\lambda }{2\pi }e^{ix\lambda }\left\langle P^{\prime
}\right| \bar{\psi}(-\frac{\lambda }{2}n,-\frac{x_{\bot }}{2})\gamma ^{\nu
_{\bot }}\psi (\frac{\lambda }{2}n,\frac{x_{\bot }}{2})\left| P\right\rangle
_{x_{\bot }=0}  \nonumber \\
&=&\bar{U}(P^{\prime },S^{\prime })\gamma ^{\nu _{\bot }}U(P,S)\frac{1}{2}%
\int dx\left[ x(H(x)+E(x))+\Delta q(x)\right] +\varpropto \Delta ^{\nu
_{\bot }}.  \label{EOM5}
\end{eqnarray}
According to the parametrization in \cite{Pol02}, this is equal to 
\begin{eqnarray*}
&&\bar{U}(P^{\prime },S^{\prime })\gamma ^{\nu _{\bot }}U(P,S)\int dxx\left[
H(x)+E(x)+G_{2}(x)\right] \\
&&-i\epsilon ^{\nu _{\bot }\beta _{\bot }np}\Delta _{\beta _{\bot }}\bar{U}%
(P^{\prime },S^{\prime })\not{n}\gamma _{5}U(P,S)\int dxxG_{4}(x)\left(
+\varpropto \Delta ^{\nu _{\bot }}\right) .
\end{eqnarray*}
Taking both equations together we get 
\begin{eqnarray}
&&\bar{U}(P^{\prime },S^{\prime })\gamma ^{\nu _{\bot }}U(P,S)\int dx\left[ -%
\frac{1}{2}x(H(x)+E(x))-xG_{2}(x)+\frac{1}{2}\Delta q(x)\right]  \nonumber \\
&=&-i\epsilon ^{\nu _{\bot }\beta _{\bot }np}\Delta _{\beta _{\bot }}\bar{U}%
(P^{\prime },S^{\prime })\not{n}\gamma _{5}U(P,S)\int dxxG_{4}(x).
\label{sumrule9}
\end{eqnarray}
which leads to the sumrule 
\[
\int dx\left[ -\frac{1}{2}x(H(x)+E(x))-xG_{2}(x)+2xG_{4}(x)+\frac{1}{2}%
\Delta q(x)\right] =0, 
\]
respectively 
\begin{equation}
\int dxx\left[ G_{2}(x)-2G_{4}(x)\right] =-J_{q}+\frac{1}{2}\Delta q=-L_{q}.
\label{sumrule11}
\end{equation}
On the other hand, 
\[
\int dxxG_{4}(x)=0, 
\]
giving the known \cite{Pen00,Pol02} result 
\[
\int dxxG_{2}(x)=-J_{q}+\frac{1}{2}\Delta q=-L_{q}. 
\]
This sumrule is gauge-invariant, and the possible contributions including
explicit gluon operators drop out for the second moment over $x$.
Furthermore, integrating (\ref{sumrule5}) over $x$ and using Ji's sumrule (%
\ref{sumrule2}) \cite{Ji97} we find 
\begin{eqnarray}
L_{q} &=&\int dxx\left( q(x)+E(x)+G_{2}(x)-2G_{4}(x)\right) -\int dx\Delta
q(x)  \nonumber \\
&=&2J_{q}-\Delta q+\int dxxG_{2}(x)  \nonumber \\
&=&2L_{q}+\int dxxG_{2}(x),  \label{sumrule12}
\end{eqnarray}
which is perfectly consistent with the sumrule eq.(\ref{sumrule11}) and
confirms our identification eq.(\ref{ident1}) on the level of integrated
distributions.

\section{Conclusions}

The quark OAM distributions in eq. (\ref{sumrule7}) and eq.(\ref
{sumrule5}) as well as the definitions in \cite{Harindranath1998,Haegler1998}
all coincide when contributions are neglected which
contain explicit transverse gluon operators $A_\perp$. 
This has been discussed in \cite{Ji99}, see eqs.(20)-(24) and the paragraph below 
eq.(28) therein.
Dropping the $A_\perp$-terms, we can therefore combine eq. (\ref{sumrule7}) and eq.(\ref
{sumrule5}) to get the interesting and simple relation 
\begin{equation}
L_{q}(x)=-x\left[ G_{2}(x)-2G_{4}(x)\right] ,  \label{sumrule8}
\end{equation}
which is obviously a generalization of the integrated sumrule eq.(\ref
{sumrule11}).

In summary we have shown that in the framework of the WW-approximation the
quark orbital angular momentum distribution is directly related to the
twist-3 GPDs $G_{2}(x)$ and $G_{4}(x),$ taken in the forward limit, in form
of the sumrule eq.(\ref{sumrule8}). Our results represent only a small step
towards solving the notorious problem of a direct measurement of the quark
OAM contribution to the nucleon spin. At least it leads to a new and nice
interpretation of the above mentioned twist-3 GPDs. It has to be seen if the
sumrule eq.(\ref{sumrule8}) or a similar expression holds outside the
WW-approximation. The main obstacle in this regard will be the use of an
accurately defined gauge invariant OAM distribution.

\subsection*{Acknowledgments}

The authors would like to thank Andreas Freund, Piet Mulders, Daniel Boer
and Fetze Pijlman for helpful discussion. Ph.H. is grateful for the support
by the Alexander von Humboldt foundation. This work has been supported by
BMBF and DFG.

\end{document}